\documentclass{pnastwo}

\usepackage{graphicx}

\usepackage{amssymb,amsfonts,amsmath}

\contributor{Submitted to Proceedings of the National Academy of Sciences of the United States of America}
\url{www.pnas.org/cgi/doi/10.1073/pnas.0709640104}
\copyrightyear{2008}
\issuedate{Issue Date}
\volume{Volume}
\issuenumber{Issue Number}

\begin{document}

%----------------------------------------------------------------------------------------
%	TITLE AND AUTHORS
%----------------------------------------------------------------------------------------

\title{Granular impact cratering by liquid drops: Understanding raindrop imprints through an analogy to asteroid strikes} % For titles, only capitalize the first letter

%------------------------------------------------

%% Enter authors via the \author command.  
%% Use \affil to define affiliations.
%% (Leave no spaces between author name and \affil command)

%% Note that the \thanks{} command has been disabled in favor of
%% a generic, reserved space for PNAS publication footnotes.

%% \author{<author name>
%% \affil{<number>}{<Institution>}} One number for each institution.
%% The same number should be used for authors that
%% are affiliated with the same institution, after the first time
%% only the number is needed, ie, \affil{number}{text}, \affil{number}{}
%% Then, before last author ...
%% \and
%% \author{<author name>
%% \affil{<number>}{}}

%% For example, assuming Garcia and Sonnery are both affiliated with
%% Universidad de Murcia:
%% \author{Roberta Graff\affil{1}{University of Cambridge, Cambridge,
%% United Kingdom},
%% Javier de Ruiz Garcia\affil{2}{Universidad de Murcia, Bioquimica y Biologia
%% Molecular, Murcia, Spain}, \and Franklin Sonnery\affil{2}{}}

\author{Runchen Zhao\affil{1}{Department of Chemical Engineering and Materials
Science, University of Minnesota, Minneapolis, MN 55455},
Qianyun Zhang\affil{1},
Hendro Tjugito\affil{1},
\and
Xiang Cheng\affil{1}{}}

\contributor{Submitted to Proceedings of the National Academy of Sciences
of the United States of America}

\significancetext{We provide a quantitative understanding of raindrop impacts on sandy surfaces---a ubiquitous phenomenon relevant to many important natural, agricultural and industrial processes. Combining high-speed photography with high-precision laser profilometry, we investigate the dynamics of liquid-drop impacts on granular surfaces and monitor the morphology of resulting impact craters. Remarkably, we discover a quantitative similarity between liquid-drop impacts and asteroid strikes in terms of both the energy scaling and the aspect ratio of their impact craters. Such a similarity inspires us to apply the idea developed in planetary sciences to liquid-drop impact cratering, which leads to a model that quantitatively describes various features of liquid-drop imprints.}

%----------------------------------------------------------------------------------------

\maketitle % The \maketitle command is necessary to build the title page

\begin{article}

%----------------------------------------------------------------------------------------
%	ABSTRACT, KEYWORDS AND ABBREVIATIONS
%----------------------------------------------------------------------------------------

\begin{abstract}
When a granular material is impacted by a sphere, its surface deforms like a liquid yet it preserves a circular crater like a solid. Although the mechanism of granular impact cratering by solid spheres is well explored, our knowledge on granular impact cratering by liquid drops is still very limited. Here, by combining high-speed photography with high-precision laser profilometry, we investigate liquid-drop impact dynamics on granular surface and monitor the morphology of resulting impact craters. Surprisingly, we find that, despite the enormous energy and length difference, granular impact cratering by liquid drops follows the same energy scaling and reproduces the same crater morphology as that of asteroid impact craters. Inspired by this similarity, we integrate the physical insight from planetary sciences, the liquid marble model from fluid mechanics and the concept of jamming transition from granular physics into a simple theoretical framework that quantitatively describes all the main features of liquid-drop imprints in granular media. Our study sheds light on the mechanisms governing raindrop impacts on granular surfaces and reveals a remarkable analogy between familiar phenomena of raining and catastrophic asteroid strikes.   
\end{abstract}

%------------------------------------------------

 % When adding keywords, separate each term with a straight line: |

%------------------------------------------------

%% Optional for entering abbreviations, separate the abbreviation from
%% its definition with a comma, separate each pair with a semicolon:
%% for example:
%% \abbreviations{SAM, self-assembled monolayer; OTS,
%% octadecyltrichlorosilane}

% \abbreviations{}

%----------------------------------------------------------------------------------------
%	PUBLICATION CONTENT
%----------------------------------------------------------------------------------------

%% The first letter of the article should be drop cap: \dropcap{} e.g.,
%\dropcap{I}n this article we study the evolution of ''almost-sharp'' fronts

\dropcap{G}ranular impact cratering by liquid drops is likely familiar to all of us who have watched raindrops splashing in a backyard or on a beach. It is directly relevant to many important natural, agricultural and industrial processes such as soil erosion \cite{1,2}, drip irrigation \cite{3}, dispersion of micro-organisms in soil \cite{4}, and spray-coating of particles and powders. The vestige of raindrop imprints in fossilized granular media has even been used to infer air density on Earth 2.7 billion years ago \cite{5}. Hence, understanding the dynamics of liquid-drop impacts on granular media and predicting the morphology of resulting impact craters are of great importance for a wide range of basic research and practical applications. 

Directly related to two long-standing problems in fluid and granular physics research, i.e., drop impact on solid/liquid surfaces \cite{6,7,8,9} and granular impact cratering by solid spheres \cite{10,11,12,13,14,15,16}, liquid drop impact on granular surfaces is surely more complicated. Although several recent experiments have been attempted to investigate the morphology of liquid-drop impact craters \cite{17,18,19,20,21}, a coherent picture for describing various features of the impact craters is still lacking. Even for the most straightforward impact-energy ($E$) dependence of the size of liquid-drop impact craters, the results remain controversial and incomplete \cite{17,19,20}. Katsuragi \cite{17} and Delon et al. \cite{19} reported that the diameter of liquid-drop impact craters, $D_c$, scales as the 1/4 power of the Weber number of liquid drops, which yields $D_c \sim E^{1/4}$, quantitatively similar to the energy scaling for low-speed solid-sphere impact cratering \cite{10,11}. However, since the energy balance of liquid-drop impacts is different from that of solid-sphere impacts, the energy scaling argument used for solid-sphere impact cratering cannot be applied to explain the 1/4 power. Instead, Katsuragi argued that the power arises from the scaling of the maximal spreading diameter of the impinging drop, which coincidently follows the same 1/4 scaling with $E$ \cite{22}. However, a later study by Nefzaoui and Skurtys showed that $D_c$ is not equal to the maximal spreading diameter and a different scaling with $D_c \sim E^{0.18}$ was found \cite{20}. Although covering a larger dynamic range of $E$, Nefzaoui and Skurtys only investigated the scaling dependence on $E$ and failed to provide a full scaling for $D_c$. It is still unclear what's the origin of the strange 0.18 scaling in liquid-drop impact cratering. Finally, in addition to the diameter of impact craters, other important properties of liquid-drop impact craters such as the depth of impact craters and the shape of granular residues inside craters have not been systematically explored so far. 

The challenges faced in the study of liquid-drop impact on granular surfaces are mainly due to the large number of relevant parameters involved in the process, the inability of existing methods for resolving the 3D structure of impact craters and the difficulty in extending the dynamic range of $E$ in experiments. Here, we investigate the dynamics of liquid-drop impacts on granular surfaces across the largest range of impact energy that has been probed so far, which covers more than four decades from the drop deposition regime to the drop terminal velocity regime. Through a systemic study using different liquid drops and granular particles at various ambient pressures, we obtain a full dimensionless scaling for the diameter of liquid-drop impact craters. Surprisingly, we find that this scaling follows the well-established Schmidt-Holsapple scaling rule associated with asteroid impact cratering \cite{23}. Moreover, by combining high-speed photography with high-precision laser profilometry, we non-intrusively measure the depth of impact craters underneath the impinging drop. The measurement reveals that liquid-drop impact craters and asteroid impact craters exhibit a self-similar shape in spite of their enormous length difference over seven orders of magnitude. These remarkable findings inspire us to apply the physical insight developed for asteroid impact cratering to the problem of liquid-drop impact cratering. The insight, in combination with the concepts of liquid marble \cite{24} and particle jamming transition \cite{25,26}, leads to a simple coherent theoretical framework that quantitatively captures all the main features of liquid-drop imprints in granular media including the diameter and the depth of impact craters and the shape of granular residues.  

\begin{figure*}
\centerline{\includegraphics[width=1.0\linewidth]{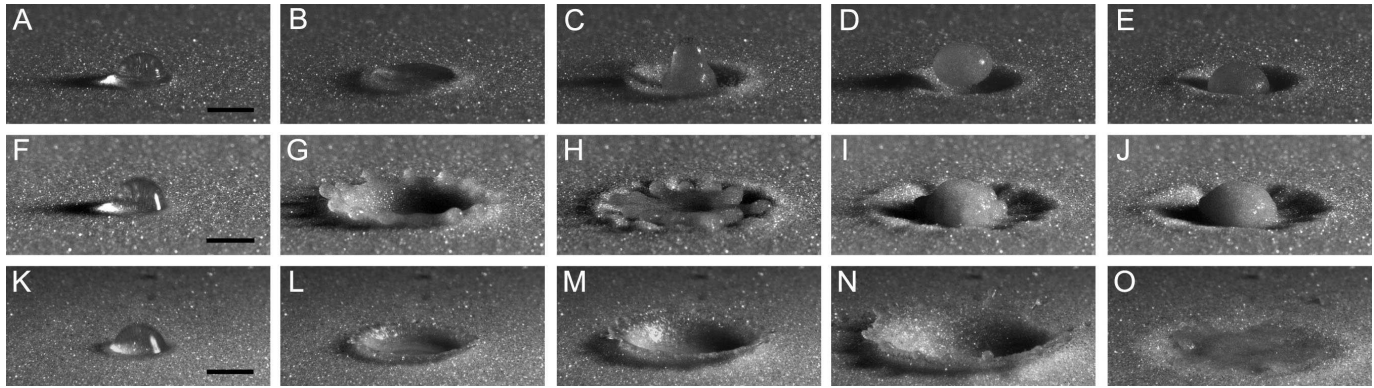}}
%select pdftexify command to run jpg or pdf files
\caption{Impact of a water drop on a granular surface. Snapshots from high-speed movies showing the impact of a 3.1 mm water drop with $E = 7.8\times10^{-6}$ J (A-E) (Movie S1), $E = 6.0\times10^{-5}$ J (F-J) (Movie S2), and $2.3\times10^{-4}$ J (K-O) (Movie S3). For the low $E$, the time elapsed after the initial impact is $t$ = 1.1 ms (A), 4.5 ms (B), 13.8 ms (C), 32.8 ms (D), and 84.0 ms (E). For the intermediate $E$, $t$ = 0.3 ms (F), 5.7 ms (G), 11.9 ms (H), 19.4 ms (I), and 56.8 ms (J).  For the high $E$, $t$ = 0.3 ms (F), 1.0 ms (G), 1.9 ms (H), 6.4 ms (I), and 29.1 ms (J). Scale bars: 3.0 mm. Water in the liquid-granular mixtures gradually drains into the bed on the time scale of a second.} 
\label{Figure1}
\end{figure*}

%------------------------------------------------

\section{Results: Liquid-drop impact dynamics}

In our experiments, we release a stationary water drop of diameter, $D$, ranging from 1.4 mm to 4.6 mm from a height $h$. $D$ is chosen to represent the size range of natural raindrops \cite{5,27}. The drop falls vertically in air onto a granular bed comprised of $d_{sand} = 90\pm15$ $\mu$m glass beads with volume fraction $\phi = 0.60$. To adjust the range of impact energy, $E$, we vary $h$ from 1.8 mm up to 12 m, allowing a 4.6 mm drop to reach $98\%$ of its terminal velocity (see Materials and Methods). 

The dynamics of liquid-drop impact on a granular surface are captured using high-speed photography as illustrated in Fig.~\ref{Figure1} for the strike of a water drop at three different $E$ (Supplementary Movie S1, S2 and S3). Upon impact, the drop penetrates into the top layer of the granular bed (Fig.~\ref{Figure1}A, F, K). After the initial impact, the drop can be treated as an incompressible fluid. The downward motion of the top part of the drop causes drop deformation and spreading.

At low $E$, the spreading liquid lamella moves horizontally along the surface of a shallow crater (Fig.~\ref{Figure1}B). The lamella retracts after reaching the maximum spreading diameter and entrains a layer of granular particles on its surface. Since the lamella's surface-to-volume ratio reduces as it recedes, particles at the interface are gradually pushed into the liquid bulk, resulting in a ``liquid marble'' armored with a thick layer of granular particles (Fig.~\ref{Figure1}C, D)\cite{24}. Above $E = 1.9\times10^{-6}$ J, the marble can even bounce off the granular bed (Fig.~\ref{Figure1}D). The jumping height of the marble is non-monotonic with increasing $E$. As the spreading diameter increases, the lamella traps more particles, which increases the weight of the marble and reduces the jumping height. 

Increasing $E$ further, the rim of the spreading lamella develops a fingering instability (Fig.~\ref{Figure1}G). After reaching the maximum spreading diameter, the fingers start to retract and gradually push particles at interface into the bulk (Fig.~\ref{Figure1}H, I). The process continues until the concentration of particles within the retracting lamella becomes so high that the receding motion is completely arrested due to the jamming of particles. The jamming transition occurs before the fingers can fully retract back to a sphere, which leads to an asymmetric liquid marble with finger protrusions on its surface (Fig.~\ref{Figure1}I, J). The length of these protrusions increases with $E$. At this intermediate $E$, the marble stops bouncing off the surface.   

At even larger $E$, a water crown is formed along the wall of a deep crater (Fig.~\ref{Figure1}L, M). The crown detaches from the granular surface at the edge of the crater. Above $E = 9.7\times10^{-5}$ J, the rim of the crown becomes unstable and disintegrates into secondary droplets (Fig.~\ref{Figure1}N). This violent splashing process effectively mixes granular particles with the liquid. Finally, the crown, fully loaded with particles, retracts and falls flat on the surface (Fig.~\ref{Figure1}O). 

The dynamics of liquid-drop impacts illustrated by high-speed photography provide essential information for understanding the morphology of liquid-drop impact craters. Based on the dynamics, we will develop a simple theoretical understanding of various features of liquid-drop impact craters in the Theory section. Before that, we shall first show our experimental results on the morphology of liquid-drop impact craters.

\section{Results: Morphology of impact craters}
\begin{figure}[ht]
\centerline{\includegraphics[width=0.95\linewidth]{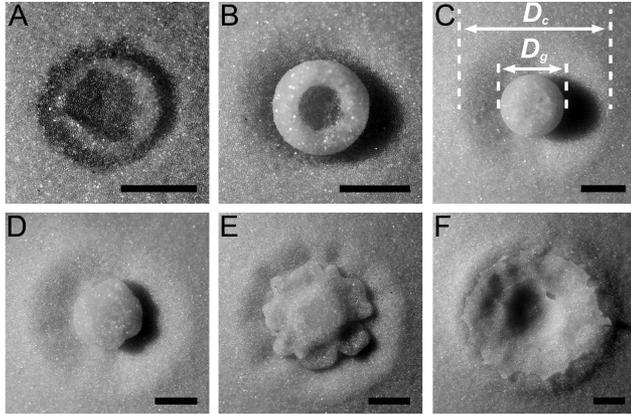}}
%select pdftexify command to run jpg or pdf files
\caption{Morphology of liquid-drop impact craters. Impact craters from the strike of a 3.1 mm water drop with $E = 9.7\times10^{-7}$ J (A), $6.5\times10^{-6}$ J (B), $3.2\times10^{-5}$ J (C), $6.0\times10^{-5}$ J (D), $8.2\times10^{-5}$ J (E) and $3.0\times10^{-4}$ J (F). Scale bars: 3.0 mm. (A) and (B) Ring-shaped granular residues at low $E$. (C) Solid pellet-shaped granular residue at intermediate $E$. (D) and (E) Asymmetric granular residues. (D) marks the transition between the low and high $E$ regime in Fig.~\ref{Figure5}. (F) Splash pattern at high $E$. $D_c$ and $D_g$ are defined in (C).} 
\label{Figure2}
\end{figure}

After impact, water gradually drains into the granular bed and various fascinating crater topologies are observed at the end (Fig.~\ref{Figure2}A-F). To fully characterize the morphology of liquid-drop imprints, we need to consider three main features of impact craters, i.e., the diameter of impact craters, the depth of impact craters and the granular residues left in the center of impact craters.

\subsection{Diameter of impact craters} 
We characterize the size of an impact crater by measuring its diameter, $D_c$ (Fig.~\ref{Figure2}C). Plotting $D_c$ versus $E$ reveals a power-law scaling with an exponent of $0.17\pm0.01$ (Fig.~\ref{Figure3}A), consistent with Nefzaoui and Skurtys's result \cite{20}. This scaling is visibly different from the 1/4 power-law scaling associated with the impact craters created by low-speed solid spheres. The 1/4 scaling of solid-sphere impacts arises when $E$ lifts granular particles of volume $\sim D_c^3$ to a height of $\sim D_c$ against the gravity \cite{10,11,23}.

Surprisingly, the 0.17 scaling is quantitatively similar to the Schmidt-Holsapple (S-H) scaling from hypervelocity impact cratering associated with asteroid strikes \cite{23}:
\begin{eqnarray}
\label{equ1} D_c & \sim & g^{-0.17}\cdot D^{0.83}\cdot U^{0.34} \nonumber \\
& \sim & (\rho g)^{-0.17}\cdot D^{0.32}\cdot E^{0.17},
\end{eqnarray}
where $U$ is the impact velocity of the projectile, $\rho$ is the density of projectile and $g$ is the gravitational acceleration. $\rho$ emerges in Eq.~\ref{equ1} when we convert $U$ into the impact energy $E$. Eq.~\ref{equ1} inspired us to apply the full S-H scaling to our data. Remarkably, we find that the variation of $D_c$ with different $D$ collapses to a constant, $C=D_c/\left((\rho g)^{-0.17}D^{0.32}E^{0.17}\right)=1.74\pm 0.15$ (Fig.~\ref{Figure3}B). Moreover, we tested the scaling using nine different liquids and seven different granular particles at two different ambient pressures. The results all conform to Eq.~\ref{equ1} (Supporting Information (SI) Fig. S1). Particularly, the $D_c$ scaling is independent of or only weakly depends on liquid properties such as density, viscosity or surface tension.

\subsection{Depth of impact craters} 

The quantitative similarity between liquid-drop impact cratering and asteroid impact cratering also extends to the aspect ratio of their impact craters, $\alpha \equiv d_c/D_c$. Here, $d_c$ is the depth of crater, defined as the vertical distance between the rim and the bottom floor of the crater.

Previous studies reported the depth of crater in the presence of granular residues \cite{17,21}, which, however, does not reflect the true bottom of a crater underneath the granular residues. Here, to detect $d_c$ without the optical obstruction of granular residues in the center of crater, we focus on the range of $E$ where the liquid marble bounces off the surface (Fig.~\ref{Figure1}D). The landing of the marble does not trigger further granular avalanche and, therefore, does not modify the crater depth at later times (Movie S1). Even though $E$ with jumping marbles does not cover the full dynamic range of our experiments, it still extends for almost two decades, allowing us to measure $d_c$ in a sufficient range comparable to other impact cratering experiments \cite{11,28}. 

Within this $E$ range, $d_c$ increases linearly with $D_c$, which leads to a constant crater aspect ratio $\alpha=0.20\pm 0.01$ (Fig.~\ref{Figure4}). As a comparison, simple craters from the Moon, Mars and Mercury also show an aspect ratio $\alpha=0.20 \pm 0.03$ \cite{29}. Even though there is a seven-order-of-magnitude difference in lengths, liquid-drop impact craters and planetary craters show the same aspect ratio within experimental errors (Fig.~\ref{Figure4}). The angle of repose of granular materials, $\theta_r$, sets an upper limit for $\alpha$. For $\theta_r = 26^\circ$---the angle measured from our experiments---we have $\alpha \lesssim \tan\theta_r/2 = 0.24$. However, the geometrical factor alone is not sufficient to explain the similarities and differences between impact cratering processes. The aspect ratio of impact craters from low-speed solid-sphere impacts is 0.12, substantially smaller than that of liquid-drop impact craters (Fig.~\ref{Figure4} inset). With strong scattering around 0.16, the aspect ratio of impact craters from hypervelocity solid-sphere impact experiments partially overlaps with that of liquid-drop impact craters (Fig.~\ref{Figure4}). 

A theoretical understanding of the scaling of the diameter and the depth of impact craters and a discussion on the similarity between liquid-drop impact cratering and asteroid impact cratering will be presented below in the Theory section.

\subsection{Granular residues}

Finally, we also measure the size of granular residues, $D_g$, in the center of impact craters (Fig.~\ref{Figure2}C). $D_g$ as a function of $E$ exhibits two different regimes (Fig.~\ref{Figure5}). At low $E$, $D_g$ slowly increases. Above certain threshold impact energy, $E^*$, it starts to enlarge strongly and merges into a master curve.

\begin{figure}[ht]
\centerline{\includegraphics[width=0.9\linewidth]{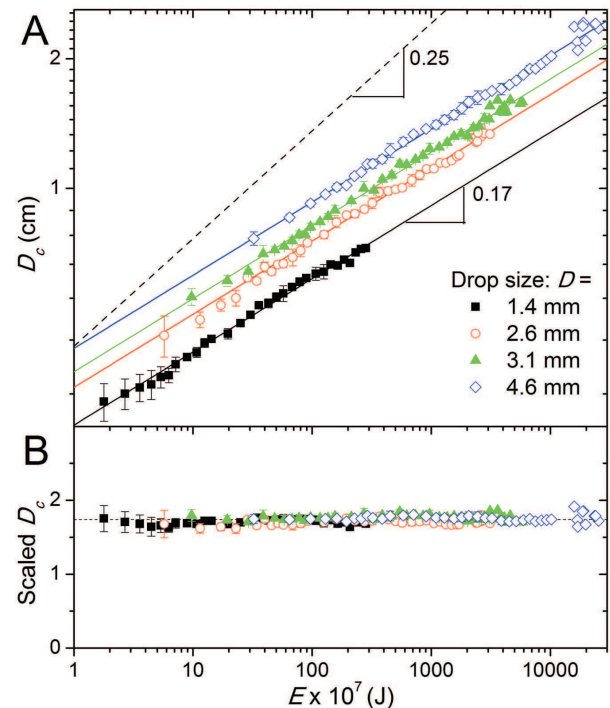}}
%select pdftexify command to run jpg or pdf files
\caption{Scaling of liquid-drop impact craters. (A) $D_c$ versus $E$ for different drop sizes. Solid lines indicate the 0.17 scaling. The dashed line indicates the 1/4 scaling. (B) Scaled $D_c$ following the S-H scaling rule (Eq.~\ref{equ1}): $D_c/((\rho g)^{-0.17}D^{0.32}E^{0.17})$. The dashed line indicates 1.74.} 
\label{Figure3}
\end{figure}

\begin{figure}[ht]
\centerline{\includegraphics[width=1.0\linewidth]{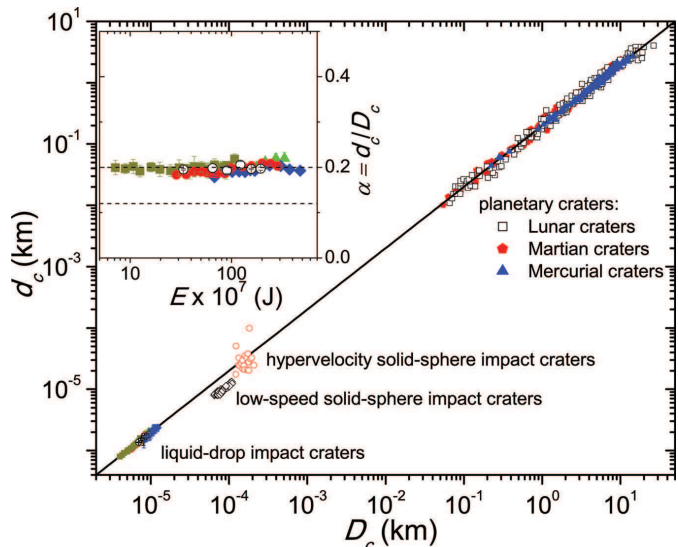}}
%select pdftexify command to run jpg or pdf files
\caption{Aspect ratio of liquid-drop impact craters. $d_c$ versus $D_c$ for four different impact cratering processes. Group (1) is from astronomical observations of asteroid impact craters on different planetary bodies \cite{29}. Group (2) is from hypervelocity solid-sphere impact experiments \cite{28}.  Group (3) is from low-speed solid-sphere impact experiments \cite{11}. Group (4) is from liquid-drop impacts with $D$ = 3.9 mm (blue diamonds), 3.1 mm (green triangles), 2.6 mm (red disks), 1.4 mm (dark yellow squares). Circles are for $D$ = 3.1 mm impacting at one tenth of the atmospheric pressure. Insets show $d_c/D_c$ of liquid-drop impact craters. The upper and lower line indicate the aspect ratio of planetary impact craters (0.20) and low-speed solid-sphere impact craters (0.12) respectively.} 
\label{Figure4}
\end{figure}  

The trend of $D_g$ can be qualitatively understood based on the impact dynamics. As shown in Fig.~\ref{Figure1}, a liquid marble coated with a layer of granular particles is formed at low $E$ during impacts. The thickness of the granular layer depends on the number of entrained particles. The liquid phase of the marble eventually drains into the granular bed and particles are left as a granular residue. With a small $E$, particles cover only the surface of the marble. Hence, when the liquid drains into the bed, a ring of particles is left (Fig.~\ref{Figure2}A). Since the maximal spreading diameter of the impinging drop increases with $E$, at larger $E$, an increasing number of particles are entrained at the lamella interface and pushed into the bulk of the liquid marble, which leads to a liquid marble with a thicker layer of granular particles. As a result, the hole at the center of the ring-shaped residues gradually fills up (Fig.~\ref{Figure2}B). At $E$ close to the transition impact energy $E^*$, particles completely saturate the marble, which leaves a solid pellet-shaped residue in the crater (Fig.~\ref{Figure2}C). Increasing $E$ above $E^*$, the receding lamella cannot fully restore back to a spherical shape due to the jamming of particles inside the marble (Fig.~\ref{Figure1}F-J), which gives rise to a flat asymmetric granular residue quickly enlarging the measurement of $D_g$ (Fig.~\ref{Figure2}D-F). A model based on the above picture will be constructed in the next section to quantitatively describe the size of granular residues.

\section{Theory and discussion}

\subsection{Understanding the S-H scaling} 

Theoretical understanding of the S-H scaling in asteroid impacts is solely based on similarity and dimensional analyses independent of detailed dynamics of impact cratering processes \cite{23,30,31}, which can thus be equally applied for liquid-drop impact cratering. However, dimensional analysis alone cannot reveal physical mechanisms associated with the liquid-drop impact process. Hence, it is more useful to look into the physical picture derived from the studies of asteroid impact cratering, which may help to explain the origin of the S-H scaling in liquid-drop impact cratering. During asteroid impacts, a large fraction of $E$ (over $97\%$ for high-velocity impacts) dissipates into heat rather than transferring to the kinetic energy of ejecta \cite{30}. The conversion efficiency of $E$ into the kinetic energy is determined by the impact pressure \cite{30,31,32}. Such a large energy partitioning is believed to give rise to the S-H scaling, which expresses a mixture of energy and momentum scaling with the power exponent between 1/4 and 1/7 \cite{30,31,32}.

Large energy partitioning also occurs during liquid-drop impacts. Only a small fraction of $E$ converts into the kinetic energy of particles, while the rest turns into the surface energy and viscous dissipation of spreading lamella. Since both the surface energy and the viscous dissipation increase with the maximal contact surface between the lamella and the granular bed ($\sim \pi D_c^2$), a larger $\pi D_c^2$ leads to a lower energy conversion. Thus, we propose a simple formula for the coefficient of energy conversion: $f = (\pi D^2/\pi D_c^2)$, where $\pi D^2$ provides the only relevant area for normalization. The fraction of energy for ejecting particles is then $E_{eject} \equiv f\cdot E$ with $f < 1$ automatically satisfied by construction. $E_{eject}$ is consistent with recent experiments that estimate the momentum of ejected particles \cite{21}. Finally, an energy scaling argument similar to that used for solid-sphere impacts can be applied: instead of $E$, $E_{eject}$ lifts granular particles in a crater of volume $V_c$ to a height determined by $d_c$, i.e., $E_{eject} \approx \phi \rho_{sand}V_cgd_c$, where $\phi = 0.60$ is the volume fraction of the bed and $\rho_{sand}$ is the particle density. If we approximate the crater as a paraboloid and replace $d_c=\alpha D_c$, then $V_c = \pi \alpha D_c^3/8$. Taken together, we successfully recover the S-H scaling:
\begin{eqnarray}
D_c \approx  \left(\frac{\pi}{8} \alpha^2\phi \frac{\rho_{sand}}{\rho}\right)^{-1/6}  [(\rho g)^{-1/6} D^{1/3}  E^{1/6}].
\label{equ2}
\end{eqnarray}
Moreover, with $\alpha = 0.20 \pm 0.01$ for liquid-drop impact craters (Fig.~\ref{Figure4}), we have the dimensionless prefactor $C \equiv \left(\frac{\pi}{8} \alpha^2\phi \frac{\rho_{sand}}{\rho}\right)^{-1/6} = 1.86 \pm 0.04$, quantitatively matching our measurement $C = 1.74 \pm 0.15$ (Fig.~\ref{Figure3}B). Hence, the scaling analysis provides a quantitative description for both the diameter and the depth of the liquid-drop impact craters.   

\begin{figure}[h]
\centerline{\includegraphics[width=0.9\linewidth]{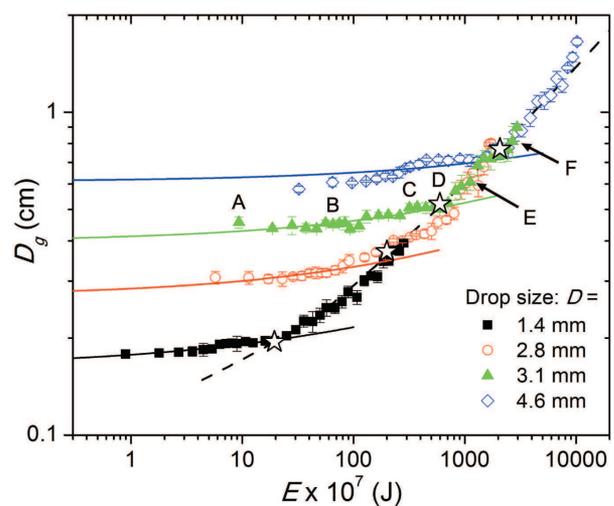}}
%select pdftexify command to run jpg or pdf files
\caption{Morphology of granular residues, $D_g$ versus $E$. Crater morphologies shown in Fig.~\ref{Figure2}A-F are indicated. Stars mark the transition impact energy $E^*$ between the low and high $E$ regime for each drop size. Solid lines are from the liquid marble model (Eq.~\ref{equ3}). The dashed line is $D_g(E^*)$ calculated by combining the liquid marble model (Eq.~\ref{equ3}) with the jamming criterion (Eq.~\ref{equ5}).} 
\label{Figure5}
\end{figure}

\subsection{Discussion on the analogy between liquid-drop impact cratering and asteroid impact cratering} 

It should be clear from the above derivation that the energy partitioning of liquid-drop impact cratering and asteroid impact cratering shares a quantitative similarity. The forms of energy dissipation in the two processes are obviously different. For asteroid impacts, the impact energy is primarily dissipated by shock-wave heating of the asteroid and the target during the initial stage of the impact event \cite{32}. For liquid-drop impacts, it dissipates mainly through the deformation and viscous dissipation of the liquid drops. However, the ratio of the energy dissipation over the total impact energy seems to follow the same quantitative trend in the two processes. Hence, it would be interesting to check if the energy conversion coefficient of asteroid impact cratering is also inversely proportional to the surface area of impact craters, i.e., $f\sim 1/D_c^2$. Without shock-wave heating or projectile deformation, a large energy partitioning does not occur in low-speed solid-sphere impact cratering. Most of the impact energy is thus directly converted into the kinetic energy of granular particles for creating impact craters, which leads to the 1/4 power as dictated by the energy scaling \cite{10,11}.   

Finally, it is also interesting to compare liquid-drop impact cratering and asteroid impact cratering more generally in terms of hydrodynamic similarity and the states of matter. Firstly, it is known that the important dimensionless number governing asteroid impact cratering is the inverse Froude number, Fr$^{-1}=gD/2U^2$ \cite{23}. For typical asteroid impacts, $10^{-6}<$ Fr$^{-1}$ $<10^{-2}$, which overlaps well with our liquid-drop impact experiments $2\times10^{-4}<$ Fr$^{-1}$ $<0.1$. Secondly, in studying asteroid impact cratering, the impacted surface is frequently modeled as a Bingham fluid \cite{32}. On the other hand, granular materials typically display Bingham fluid behavior \cite{33}. More importantly, during asteroid strikes, the impact pressure can rise as high as $10^3$ GPa and the temperature may increase above 2000 $^\circ$C. Under such extreme conditions, asteroids of normal composition have already been liquefied if not vaporized \cite{32}. Hence, liquid drops provide a better model than solid spheres for high-energy asteroids. This important analogy has been overlooked in many previous attempts in search of the link between asteroid impact cratering and low-speed solid-spheres impact cratering \cite{11,12,15,16,34,35}. 

\subsection{Model for granular residues} 

The model for granular residues can be divided into two parts: (1) Based on the liquid marble model, we will show a quantitative understanding of the size of granular residues at low $E$. (2) Employing the concept of the jamming transition, we will calculate the transition energy $E^*$ between the low and high energy regimes (Fig.~\ref{Figure5}).

(1) As shown previously, the slow increase of $D_g$ at low $E$ is due to the formation of liquid marbles (Fig.~\ref{Figure2}A-C). $D_g$ in this regime is equal to the diameter of liquid marbles. A simple model can thus be constructed based on the liquid marble model proposed by  Aussillous and Qu{\'e}r{\'e} \cite{24}. Firstly, the number of entrained particles at the lamella-granular bed interface, $N$, is proportional to the maximal contact area between the lamella and the bed. Therefore, $N\approx (\pi D_c^2/\pi d_{sand}^2)$. The volume of the liquid marble is simply the sum of the volume of the drop and the volume of entrained particles: $V_{m}=V_{drop}+V_{sand}=\pi D^3/6+N\pi d_{sand}^3/6=\pi D^3/6+\pi d_{sand}D_c^2/6$. If we assume the marble is spherical, then the effective diameter of the liquid marble is $D_m = (6V_m/\pi)^{1/3}$. For $D_m \ll \kappa^{-1}$, the liquid marble maintains a spherical shape, where $\kappa^{-1} = (\gamma/\rho_m g)^{1/2}$ is the capillary length, $\gamma$ is the surface tension of the liquid and $\rho_m$ is the density of the liquid-granular mixture. The diameter of the liquid marble and, therefore, the diameter of the granular residue is simply $D_g = D_m = (6V_m/\pi)^{1/3}$. However, for $D_m \gg \kappa^{-1}$, the marble deforms into a puddle under the force of gravity. The thickness of the puddle is given by $2\kappa^{-1}$. If we approximate the shape of the puddle as an oblate ellipsoid, then the diameter of the marble is given by $D_g = (3V_m/\pi\kappa^{-1})^{1/2}$. In summary, we have 
\begin{equation}
 D_g = \left\{
\begin{array}{rl}
C_1\cdot(6V_m/\pi)^{1/3} & \text{if } D_m \ll \kappa^{-1},\\
C_2\cdot(3V_m/\pi\kappa^{-1})^{1/2} & \text{if } D_m \gg \kappa^{-1}.
\end{array} \right.
\label{equ3}
\end{equation}
where we add two proportionality constants $C_1$ and $C_2$ to account for the fact that $D_m$ is close to $\kappa^{-1}$ between the two limiting cases and the approximation taken for the shape of the puddle. Replacing $D_c$ in $V_m$ using the S-H scaling (Eq.~\ref{equ2}), we finally reach $D_g(E)$. The results quantitatively agrees with our measurements (solid lines in Fig.~\ref{Figure5}) with the fitting parameters $C_1 = 1.1$ and $C_2 = 1.55 \pm 0.15$ on the order of one.

(2) Increasing $E$ further, at the transition impact energy $E^*$, the retraction of lamella is arrested before it can fully restore back to a sphere, which leads to asymmetric granular residues with quickly enlarging $D_g$ and results in a crossover from the low-energy ``liquid marble'' regime to the high-energy regime (Fig.~\ref{Figure5}). As discussed previously, the resistance against the capillary retraction comes from the jamming of entrained particles. Thus, we can identify $E^*$ as the ``jamming energy'' of liquid-drop impact process. Note that the particles entrained at the liquid interface are gradually pushed into the interior of the receding liquid lamella due to the strong capillary retraction. Hence, the jamming occurs in the bulk of liquid marble rather than only at its interface \cite{36}. 

\begin{figure}[h]
\centerline{\includegraphics[width=0.7\linewidth]{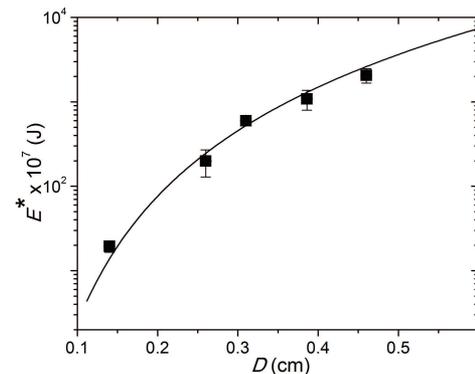}}
%select pdftexify command to run jpg or pdf files
\caption{Transition energy $E^*$ versus drop size $D$. Solid line is based on the jamming criterion (Eq.~\ref{equ5}).} 
\label{Figure6}
\end{figure}

A simple analysis based on the liquid marble model can show that the number of entrained particles at the lamella-granular interface is not sufficient to jam the liquid marble at $E^*$. To reach the jamming transition, the effect of liquid imbibition during the impact needs to be considered \cite{19}. We estimate the volume of imbibed liquid into the bed, $V_{imb}$, based on the well-established Washburn-Lucas equation \cite{37}, which leads to the following equation specifically for liquid-drop impact cratering (SI): 
\begin{eqnarray}
V_{imb}=0.058A\cdot (\eta^2\gamma)^{-1/4}\cdot \rho^{1/4}d_{sand}\cdot D^{5/4}\cdot E^{1/2},
\label{equ4}
\end{eqnarray}
where $\eta$ is the liquid viscosity and $A$ is a proportionality constant of order one. The jamming transition at $E^*$ can then be expressed as: 
\begin{eqnarray}
\frac{V_{sand}}{V_m-V_{imb}} =  \frac{\pi d_{sand}D_c^2/6}{\pi D^3/6 +\pi d_{sand}D_c^2/6 - V_{imb}}  =  \phi_c
\label{equ5}
\end{eqnarray}
with the jamming volume fraction $\phi_c \approx 0.55$ \cite{25,26}. Using the S-H scaling for $D_c$, Eq.~\ref{equ5} quantitatively agrees with our measurement on $E^*$ for different $D$ with the fitting constant $A = 1.19 \pm 0.22$ (Fig.~\ref{Figure6}). Finally, in combination with the liquid marble model, we also reach $D_g(E^*)$ (dashed line in Fig.~\ref{Figure5}) (see SI for additional comments).

\section{Conclusions}

When a liquid drop impacts on a granular surface, the impact energy is converted into the surface energy of the deformed drop, the internal energy of liquid and particles, and the kinetic energy of the spreading lamella and ejected particles. The process is notoriously complicated, involving high Reynolds hydrodynamics, shock compression in the impinging drop, fast granular flows and capillary interactions between fluid and granular particles. Given the complexity, it is surprising that the simple model presented here can quantitatively captures the morphology of liquid-drop impact craters over a large range of impact energy. Such a model will be considerable useful for predicting the outcome of raindrop impacts on granular media---a ubiquitous process occurring in numerous natural, agricultural and industrial circumstances. 

Moreover, our study reveals a quantitative similarity between raindrop impact cratering and asteroid impact cratering in terms of both the energy scaling and the aspect ratio of their impact craters. Comparing with extensively-studied low-speed solid-sphere impact cratering, liquid-drop impact cratering provides a better analogy to high-energy asteroid impact cratering. Apparently, one should be very cautious when drawing a close link between the two processes. $E$ of an asteroid is on the order of $10^{15}$ J \cite{32}, while the maximal $E$ of liquid drops is $10^{-3}$ J. The 18-order-of-magnitude energy difference undoubtedly activates different physical processes. Nevertheless, the remarkable similarity between the two processes indicates that they may share common mechanisms that are worth further investigation. 

\begin{materials}
For all the experiments and data presented in the main text, we used deionized water as our liquid drops and 90 $\pm$ 15 $\mu$m soda-lime glass beads ($\rho = $2.52 g/cm$^3$, MoSci) as our granular particles. In Supporting Information (SI), we also tested glass particles of different sizes (45.5 $\pm$ 7.5 $\mu$m, 215 $\pm$ 35 $\mu$m, 427 $\pm$ 73 $\mu$m, 725 $\pm$ 125 $\mu$m) and wetting properties to verify the S-H scaling. Moreover, in SI, we used several different liquids including methanol, ethylene glycol, mineral oil, water-glycerin mixtures and sodium dodecyl sulfate solutions as liquid drops to probe the effect of liquid viscosity and surface tension on the impact cratering process. A brief discussion of liquid-drop impact cratering on wet granular bed is also presented in SI.   

A Photron SA-X2 camera was used for high-speed imaging of drop impact dynamics. The morphology of impact craters were measured using a high-precision laser profilometer (Kenyence LJ-V7060) with the resolution in the x-y plane at 20 $\mu$m and the resolution in depth at 0.4 $\mu$m. The camera and the profilometer was further combined to monitor the depth of crater during impacts. Experiments in the terminal velocity regime were conducted in an indoor laboratory with a high-height experimental platform. To prevent perturbation from air flows that lead to uncontrollable impact positions, we set up a PVC tube of 11.5 m in length and 20 cm in diameter. A free falling drop travels inside the tube before it impacts on a granular bed underneath the bottom opening of the tube. The release heights in previous investigations are all below 3 m \cite{17,18,19,20,21}, which seriously limits the dynamic range of impact energy and thus the accuracy of the scaling relationship. Finally, we also performed one set of experiments at one tenth of the atmospheric pressure to test possible effects of ambient air on the dynamics of liquid-drop impact cratering. The ambient air has been shown to play a significant role in liquid-drop impacts on solid surfaces  \cite{8,9}.
\end{materials}

%----------------------------------------------------------------------------------------
%	ACKNOWLEDGEMENTS
%----------------------------------------------------------------------------------------

\begin{acknowledgments}
We thank J. Hong and W. Suszynski for the help with experiments and J. Melosh for the comments on asteroid impact cratering. We also thank F. Bates, K. Dorfman, L. Francis, S. Kumar, L. Xu and L.-N. Zou for suggestions on the paper. R.Z and Q.Z. acknowledge support from UMN UROP program.
\end{acknowledgments}

%----------------------------------------------------------------------------------------
%	BIBLIOGRAPHY
%----------------------------------------------------------------------------------------

%% PNAS does not support submission of supporting .tex files such as BibTeX.
%% Instead all references must be included in the article .tex document. 
%% If you currently use BibTeX, your bibliography is formed because the 
%% command \verb+\bibliography{}+ brings the <filename>.bbl file into your
%% .tex document. To conform to PNAS requirements, copy the reference listings
%% from your .bbl file and add them to the article .tex file, using the
%% bibliography environment described above.  

%%  Contact pnas@nas.edu if you need assistance with your
%%  bibliography.

% Sample bibliography item in PNAS format:
%% \bibitem{in-text reference} comma-separated author names up to 5,
%% for more than 5 authors use first author last name et al. (year published)
%% article title  {\it Journal Name} volume #: start page-end page.
%% ie,
% \bibitem{Neuhaus} Neuhaus J-M, Sitcher L, Meins F, Jr, Boller T (1991) 
% A short C-terminal sequence is necessary and sufficient for the
% targeting of chitinases to the plant vacuole. 
% {\it Proc Natl Acad Sci USA} 88:10362-10366.

%% Enter the largest bibliography number in the facing curly brackets
%% following \begin{thebibliography}

%----------------------------------------------------------------------------------------

\end{article}

\end{document}